# Discussion of Loop Expansion and Introduction of Series Cutting Functions to Local Potential Approximation: Complexity Analysis Using Green's Functions, Cutting Of Nth-Order Social Interactions For Progressive Safety

Yasuko Kawahata [†]

Faculty of Sociology, Department of Media Sociology, Rikkyo University, 3-34-1 Nishi-Ikebukuro,Toshima-ku, Tokyo, 171-8501, JAPAN.

ykawahata@rikkyo.ac.jp

**Abstract:** In this study, we focus on the aforementioned paper, "Examination Kubo-Matsubara Green's Function Of The Edwards-Anderson Model: Extreme Value Information Flow Of Nth-Order Interpolated Extrapolation Of Zero Phenomena Using The Replica Method(2024)". This paper also applies theoretical physics methods to better understand the filter bubble phenomenon, focusing in particular on loop expansions and truncation functions. Using the loop expansion method, the complexity of social interactions during the occurrence of filter bubbles will be discussed in order to introduce series, express mathematically, and evaluate the impact of these interactions. We analyze the interactions between agents and their time evolution using a variety of Green's functions, including delayed Green's functions, advanced Green's functions, and causal Green's functions, to capture the dynamic response of the system through local potential approximations. In addition, we apply truncation functions and truncation techniques to ensure incremental safety and evaluate the long-term stability of the system. This approach will enable a better understanding of the mechanisms of filter bubble generation and dissolution, and discuss insights into their prevention and management. This research explores the possibilities of applying theoretical physics frameworks to social science problems and examines methods for analyzing the complex dynamics of information flow and opinion formation in digital society.

## 1. Introduction

This study theoretically explores the filter bubble phenomenon and replica symmetry breaking in digital society using the Spinglass model. The filter bubble phenomenon refers to the bias in information flow caused by the emphasis of only certain information or opinions, and has an important impact on opinion formation in society. This phenomenon refers to a situation in which only certain opinions or information is emphasized due to information bias or echo chamber effect, and its impact on information distribution and opinion formation is an extremely important issue in modern society. The purpose of this study is to address this issue by analyzing correlations between multiple copies (replicas) of a system, calculating free energies, and examining replica symmetry breaking in detail.

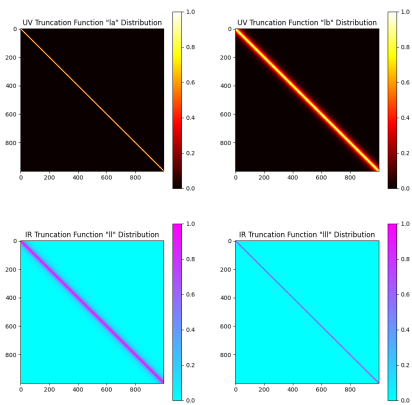

Fig. 1: Ex:Distribution of ultraviolet (UV) and infrared (IR) cutoff functions by "N"th Order

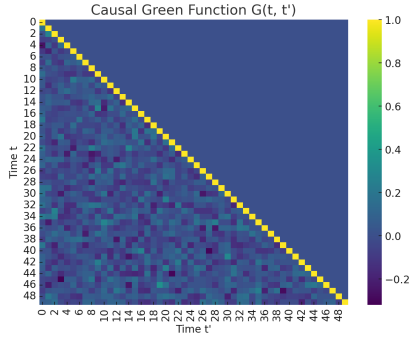

Fig. 2: Ex:Causal Green Function,Nth-Order Interpolated Extrapolation of Zero Phenomena

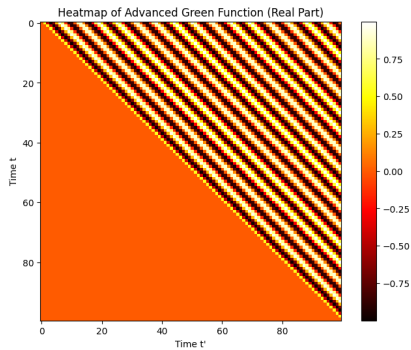

Fig. 3: Ex:Advanced Green Function,Nth-Order Interpolated Extrapolation of Zero Phenomena

### Positioning of Spin Glass Theory and the Edwards-Anderson Model

Spin glass theory is a branch of physics that analyzes the behavior of disordered magnetic materials. The theory describes the interactions between randomly oriented magnetic spins and contributes to a deeper understanding of many physical phenomena. The Edwards-Anderson model is one of the core models of spin glass theory, which describes the interaction between randomly oriented spins in a simple form. This model captures the fundamental properties of spin glass and provides a basic framework for studying disordered systems.

### Theoretical Exploration of the Filter Bubble Phenomenon

In modern society, filter bubble phenomena caused by information bias and echo chamber effects have a significant impact on opinion formation and information distribution. This study uses the Edwards-Anderson model to theoretically analyze replica symmetry breaking in this filter bubble phenomenon. Replica symmetry breaking, which occurs when considering disorder and uncertainty in the system, is key to understanding the mechanisms of filter bubble formation and dissolution.

### Design of Computational Experiments

In this study, we will use a spin glass model that includes both remote and proximity interactions, with a particular focus on the long-run average behavior in the $n$ order extrapolation and the $n$ order interpolation states during zero events. To accomplish this, we will perform a correlation analysis of these states using the replica method and consider their extreme values. This allows us to analyze in more detail the impact of information distribution and opinion formation on the formation of filter bubbles, and to deepen our theoretical understanding of the mechanisms of information bias and echo chamber effects, as discussed in the aforementioned paper (Examination Kubo-Matsubara Green's Function Of The Edwards-Anderson Model: Extreme Value Information Flow Of Nth-Order Interpolated Extrapolation Of Zero Phenomena Using The Replica The main objective of this study is to examine the social and economic aspects of the model.

In particular, this paper is about expressing the complexity of social interactions using a mathematical technique called loop expansion and quantitatively evaluating its impact. Loop expansion is a technique used in quantum field theory to deal with interactions between particles, and here it is applied to a social science problem. This method allows loops of different orders to illustrate the complexity of social interactions, and each considers the direct and indirect effects of the method. In addition, a technique called "truncation" is used to manage computational complexity. Truncation is a technique that truncates certain terms from an infinite-dimensional problem in order to allow the computational problem to be handled in a finite dimension. We hypothesize that this reduces the complexity of the model and facilitates computation, while allowing us to evaluate the impact of the model on the physical behavior of the model. Building on this argument, we use the Green function to analyze the interactions between agents in more detail and capture the dynamic response of the system through local potential approximations. The Green's function indicates the strength of the impact of an agent's state on other agents, and based on this, we aim to discuss the strength and direction of the interaction at each point in time with respect to the direction of the evaluation.

## 2. Discussion:Introducing Local Potential

The following hypotheses regarding the computational process of filter bubbles in this paper and their novelty may be considered. Quantify the complexity of social interactions using the Spinglass model to gain a deeper understanding of the filter bubble phenomenon. By mathematically expressing irregularities and promiscuity in social opinion formation through the application of spin state ordering and the Edwards-Anderson model, and by analyzing replica symmetry breaking in the filter bubble phenomenon from the perspective of possible replica symmetry breaking analysis by applying the properties of spin glass, Clarify the mechanism of how information bias and echo chamber effects affect opinion formation. This will enable us to understand the process of filter bubble formation and collapse. And further applications of loop expansion and truncation: Loop expansion and truncation techniques will be used to mathematically represent the complexity of social interactions while managing computational complexity. This will allow us to seek a detailed analysis of social interactions in information flow and opinion formation through a detailed analysis of interactions using Green's functions. These Green functions will be used to analyze the interactions between agents in detail to capture the dynamic response of the system. This will allow us to consider a more specific discussion of the intensity and direction of interactions over time.

Novelty of this study Provides a different perspective on the concept of information bias and opinion formation. Understanding information bias and opinion formation in filter bubbles and their formation process through the Spinglass model has not been explored in previous studies.

This paper describes the idea of introducing local potential approximations, loop expansions, truncation techniques, and methods for calculating minima and maxima.

Significance of Introducing Local Potential Approximation Local Potential Approximation is a method of approximating the potential (in this case the state of opinion or information) at each point in a system. This approximation allows one to model how the opinions of each agent (individual or group) in the system affect the other agents. In the context of filter bubbles, this approximation can be used to examine how information bias and echo chamber effects affect the opinion formation of individual agents.

Significance of Introducing Loop Unfolding Loop expansion is a method of series expansion of interactions in a system. Through this expansion, more complex interactions (e.g., indirect effects and long-range interactions) can be better understood. In filter bubbles, this method allows for a more detailed analysis of the flow of information and the process of opinion formation, and scrutinizes the exploration of how this can lead to bias and bias.

Significance of Introducing the Cutting Function (la,lb,ll,lll) The truncation function (la,lb,ll,lll,llll) is a way to truncate some terms from an infinite-dimensional problem to manage computational complexity. In analyzing filter bubbles, truncation allows one to focus on the most important elements (e.g., the strongest interactions or most influential agents) within the computable range. This holds the promise of efficiently handling the core aspects of the filter bubble phenomenon.

Significance of the introduction of the calculation of minima and maxima The calculation of minima and maxima is important for understanding system stability and instability. In a filter bubble, these values may indicate points of agreement or division. Minima may represent stable states of opinion, while maxima may represent instability or turning points of opinion. This analysis holds the promise of understanding the conditions under which filter bubbles form and the scenarios under which they may burst.

## 3. Discussion:Approaches to Causal Green Functions in Spin Glass Models

In the second discussion, we will discuss the causal Green's functions, advanced Green's functions, and delayed Green's functions mentioned in the aforementioned paper, as well as the We will organize the discussion regarding the solution method on how to approach the filter bubble state using the spin glass in this paper.

### Definition of Causal Green Function

The causal Green function $G(t, t')$ shows how the state of the system at time $t$ is influenced by the state at past time $t'$. In the spin glass model, this is used to represent the time-dependent interactions between spins.

$$G(t, t') = -i\langle T[S_i(t)S_j(t')]\rangle$$

Here,

$S_i(t)$ represents the state of spin $i$ at time $t$,

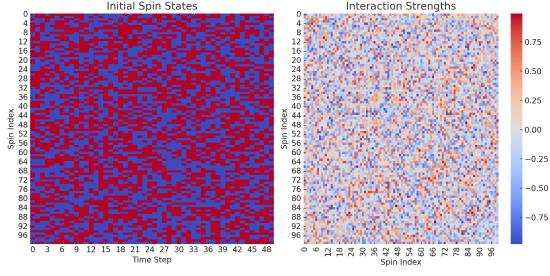

Fig. 4: Spin State, Causal Green Function, Nth-Order Interpolated Extrapolation of Zero Phenomena

$T$ is the time-ordering operator,

$\langle \cdot \rangle$ denotes the thermal average.

## Dynamics of the Spin Glass Model

The dynamics in the spin glass model are defined including the time-dependent interactions between agents. The Hamiltonian is as follows:

$$H = -\sum_{\langle i,j \rangle} J_{ij}(t) S_i(t) S_j(t)$$

Here, $J_{ij}(t)$ represents the strength of interaction between spins $i$ and $j$ at time $t$.

## Breaking of Replica Symmetry

The analysis of the breaking of replica symmetry requires the computation of the free energy, which includes correlations between different replicas. This is done based on the dynamics of interactions represented using causal Green functions.

$$F = -\frac{1}{\beta} \lim_{n \to 0} \frac{\log[Z^n]_{av}}{n}$$

Here, $[Z^n]_{av}$ represents the ensemble average of the partition function $Z^n$ based on the probability distribution of interactions.

## Behavior of Long-Time Averages

The analysis of the behavior of long-time averages involves the calculation of time evolution using causal Green functions. This allows tracking the time-dependent dynamics of the formation and dissolution of filter bubbles.

$$\langle S_i(t) \rangle = \int dt' G(t, t') \langle S_i(t') \rangle$$

We will discuss the change in the causal Green's function and the change in the number of replicas in the $n$ order extrapolation during zero phenomena. Zero phenomenon refers to the breaking of replica symmetry in spin glass theory and is related to analyzing the behavior in the limit where the number of replicas $n$ approaches zero. A filter bubble refers to a situation in which only certain information or opinions are emphasized, and in the spin glass model, the interaction between spins can change with time to mimic a situation in which some opinions become dominant.

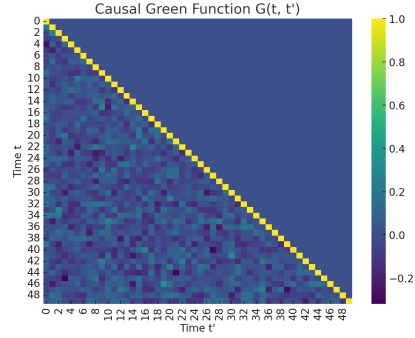

Fig. 5: Causal Green Function, Nth-Order Interpolated Extrapolation of Zero Phenomena

## Initial spin state and interaction strength

The image on the left shows the initial spin state, with spins randomly pointing upward (red) or downward (blue). This randomness represents the disorder of the spin glass state and how it evolves over time. The image on the right shows the strength of the interactions between spins, with red representing positive interactions (forces of attraction) and blue representing negative interactions (forces of repulsion). The pattern of interactions changes with time, which determines the dynamics of the spin state.

## Variation of the Causal Green's Function

The heat map of the causal Green's function shows how the spin state at time $t$ is affected by the state at past time $t'$. The heatmap is symmetric with respect to the time difference $t - t'$, showing a maximum (yellow) at $t = t'$ and a decreasing influence (blue) as the time difference increases.

## Change in number of replicas and filter bubble

Changes in the number of replicas $n$ are important in the calculation of the Green's function. By varying $n$, we can analyze how the interaction between spins depends on time and how it affects the energy state and order parameters of the system. When a filter bubble occurs, the interaction between certain spins may be enhanced while other interactions are suppressed. This is analogous to the social phenomenon of extreme polarization of opinion, and the Spinglass model can be used to study such social dynamics.

# 4. Discussion: Introduction to Advanced Green Functions in Spin Glass Models

In this discussion, we will provide a detailed explanation of the calculation process when introducing advanced Green functions into the analysis of the spin glass model.

## Definition of Advanced Green Function

The advanced Green function $G^{\text{adv}}(t, t')$ describes how the response of the physical system evolves from time $t'$ to time $t$. It is usually expressed as follows:

$$G^{\text{adv}}(t, t') = -i\theta(t' - t)\langle\{S_i(t), S_j(t')\}\rangle$$

Here,

$\theta(t)$ is the Heaviside step function,

$\{S_i(t), S_j(t')\}$ is the time $t$ and $t'$ anti-commutator of spins,

$\langle\cdot\rangle$ denotes the thermal average.

## Dynamics of the Spin Glass Model

The Hamiltonian of the spin glass model is defined to include interactions between spins, but by introducing advanced Green functions, the time-dependence of these interactions is taken into account.

$$H(t) = -\sum_{\langle i,j \rangle} J_{ij}(t) S_i(t) S_j(t)$$

## Breaking of Replica Symmetry

The breaking of replica symmetry is analyzed using the replica method. This method considers correlations between different replicas and can capture the system's non-equilibrium states.

(1) **Application of the Replica Method**: Consider $n$ replicas and calculate the partition function $Z^n$.

(2) **Analysis of Breaking of Replica Symmetry**: Analyze the breaking of replica symmetry from the results obtained using advanced Green functions.

## Behavior of Long-Time Averages

The behavior of long-time averages is obtained by analyzing how the system evolves with time using advanced Green functions.

$$\langle S_i(t) \rangle = \int dt' G^{\text{adv}}(t, t') \langle S_i(t') \rangle$$

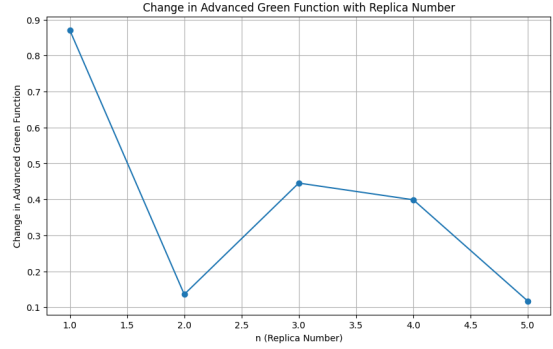

Fig. 6: Change in number of replicas, Advanced Green Function, Nth-Order Interpolated Extrapolation of Zero Phenomena

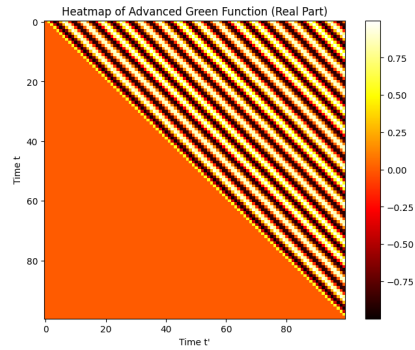

Fig. 7: Advanced Green Function, Nth-Order Interpolated Extrapolation of Zero Phenomena

### Relationship between the change in the advanced Green's function and the number of replicas

The first graph shows the change in the advanced Green's function with respect to the number of replicas $n$. The behavior of the change in the graph suggests that the Green's function is changing non-monotonically as the number of replicas increases. This may be related to replica symmetry breaking. In spin glass theory, when considering the limit of $n \to 0$, such a non-monotonic change is considered a sign of replica symmetry breaking.

### Heat Map of Advanced Green's Function

The second heat map shows the real part of the advanced Green's function $G^{\text{adv}}(t, t')$ as a function of time, where a particular pattern emerges. This pattern may represent a temporal change in the interactions between spins and the associated evolution of the spin states. In particular, the heatmap may indicate the periodicity and regularity of interactions in the filter bubble phenomenon. A filter bubble refers to a phenomenon in which information or opinions are restricted by a particular pattern or bias, which can be mimicked through patterns of strengthening or weakening of interactions in a spin-glass model.

### Considerations During Filter Bubble Occurrence

During filter bubble outbreaks, interactions between specific spins are prioritized and a state is formed in which the system is dominated by those interactions. This can also be seen in social contexts as a phenomenon in which certain opinions and information are emphasized over others. The periodicity exhibited by the pattern of advanced Green's functions may suggest how such filter bubbles are generated and maintained over time.

## 5. Discussion:Introduction of Delay Green Functions in Spin Glass Analysis

In this section, we will provide a detailed explanation, with the use of equations, of the specific calculation process when introducing delay Green functions in the analysis of spin glasses.

The delay Green function $G^{\text{ret}}(t, t')$ is defined as follows:

$$G^{\text{ret}}(t, t') = -i\theta(t - t')\langle[S_i(t), S_j(t')]\rangle$$

Here,

$\theta(t)$ is the Heaviside step function,

$[S_i(t), S_j(t')]$ is the commutator of spins at times $t$ and $t'$,

$\langle \cdot \rangle$ denotes thermal averaging.

### Dynamics of the Spin Glass Model

The Hamiltonian of the spin glass model is expressed as follows:

$$H(t) = -\sum_{\langle i,j \rangle} J_{ij}(t) S_i(t) S_j(t)$$

### Breaking Replica Symmetry

Breaking of replica symmetry can be captured by analyzing correlations between different replicas.

### Application of Replica Method

Apply delay Green functions to $n$ replicas and calculate the partition function $Z^n$.

### Analysis of Replica Symmetry Breaking

Analyze the breaking of replica symmetry based on the dynamics using delay Green functions.

### Behavior of Long-Time Averages

The behavior of long-time averages is analyzed using delay Green functions.

### Calculation of Time Evolution

Calculate the time evolution of the system based on delay Green functions.

$$\langle S_i(t) \rangle = \int dt' G^{\text{ret}}(t, t') \langle S_i(t') \rangle$$

The delayed Green's function $G^{\text{ret}}(t, t')$ is used to describe how a physical system responds to past perturbations. This function shows how the response of the system at time $t$ depends on perturbations in the past time $t'$. Specifically, $G^{\text{ret}}(t, t')$ is nonzero only at $t > t'$ and zero at $t < t'$ because it represents the causal response of the system.

The heatmap provided shows how the magnitude of the delayed Green's function changes as a function of time, with a noticeable change at certain time steps. This suggests that the spin dynamics in the spin glass model is non-uniform as a function of time. In particular, the nonzero values of $G^{\text{ret}}(t, t')$ are directly related to the strength of the interaction between $t'$ and $t$, which captures how this interaction evolves with time.

As for the relationship between the number of replicas $n$ and the advanced Green's function, this is important when analyzing the behavior of the system when replica symmetry breaking is considered. Replica symmetry breaking plays a central role in understanding the behavior of spin glass phase transitions and order parameters. The development of filter bubbles represents a social phenomenon in which information and opinions are biased toward certain groups or situations, which in spin glass models can be mimicked by the strengthening or weakening of interactions between certain spins over time.

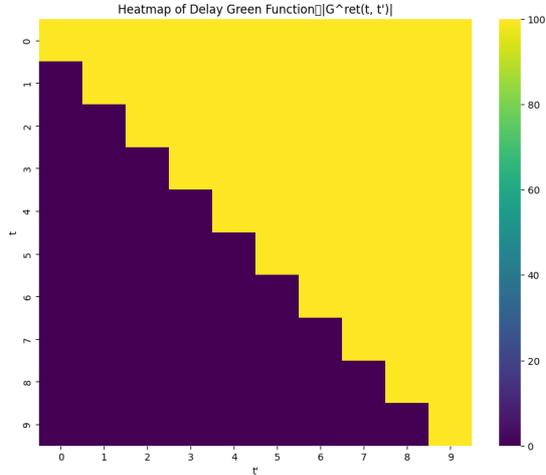

Fig. 8: Heatmap of Delay Green Function, Nth-Order Interpolated Extrapolation of Zero Phenomena

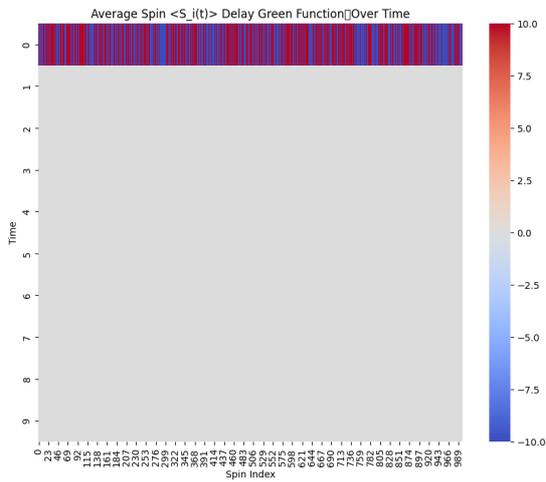

Fig. 9: Average Spin <$S_i(t)$> Delay Green Function Over Time, Nth-Order Interpolated Extrapolation of Zero Phenomena

We also use heatmaps of the delayed Green's function to provide a perspective on the onset of filter bubbles.

A filter bubble is a phenomenon in which certain information or opinions become dominant and suppress others. This creates a situation in society where individuals are exposed only to information that is consistent with their preferences and opinions, which can lead to a loss of diversity of opinion and reinforcement of extreme views and prejudices.

Insights from the Lagging Green Function Heat Map Light colors (yellow) on the heat map indicate strong interactions or strong responses, while dark colors (purple) indicate weak interactions or weak responses. The stepwise pattern shown by the heatmap may indicate that certain interactions strengthen and others diminish over time. This pattern can provide insight into how filter bubbles form over time and what their dynamics are.

### Effects of Filter Bubbles

In the spin glass model, the effect of filter bubbles can occur when some spin-to-spin interactions become dominant on certain time scales. This means that some spins interact more strongly than others, creating states that dominate the dynamics of the system. As a result, the system may converge to a particular ordered state and diverse states are suppressed.

### Application to Social Contexts

In social contexts, filter bubbles appear as a phenomenon in which individuals are surrounded by similar opinions and information, reinforcing a particular point of view. When the Spinglass model is applied to the study of social dynamics, the lagged Green's function can be used to model how an individual's opinions are shaped over time by social interactions. Mean-spin heat maps can show how opinions homogenize or lose diversity over time.

### Conclusion

Heatmaps of the lagged Green's function provide an important tool for understanding the dynamics of filter bubbles using the spin glass model. This provides insight into how information is filtered and how certain interactions become dominant over time. These models provide a powerful analogy for studying social opinion dynamics and could be connected to understanding information distribution, opinion formation, and group decision-making processes.

# 6. Discussion: Consideration of Renormalization Group in the Context of Ultraviolet and Infrared Critical Phases

When seeking local approximations in the context of zero phenomenon, n-th order extrapolation, and interpolation, it is important to consider the divergence and convergence of the renormalization group (RG) from the perspectives of ultraviolet critical phases (UV) and infrared critical phases (IR). This is particularly relevant in the analysis of complex systems in statistical physics, such as spin glass models. Below, we explain the theoretical framework for this.

### Ultraviolet Critical Phase (UV)

In the ultraviolet critical phase, we consider the behavior of the system at small scales. This includes characteristics of high-energy particle interactions and short-range interactions.

### Infrared Critical Phase (IR)

In the infrared critical phase, we consider the behavior of the system at large scales. This focuses on long-range interactions and the dynamics of the overall system at low energies.

### Divergence and Convergence of Renormalization Group

Divergence implies that the flow of the renormalization group tends toward infinity. This may indicate that specific parameters of the system become infinitely large at high-energy scales in the ultraviolet critical phase. This phenomenon can reveal non-physical properties of the theory and requires proper renormalization.

### Understanding Convergence

Convergence means that the renormalization group flow approaches specific fixed points. This indicates that the system converges to certain physical behaviors at long-range scales in the infrared critical phase. The behavior of the system near fixed points often exhibits universal characteristics shared across different models.

### Applications in Statistical Physics

In statistical physics, the theory of renormalization group is employed to understand phase transitions and critical phenomena. When seeking local approximations in the context of zero phenomenon, n-th order extrapolation, and interpolation, considering the perspectives of ultraviolet and infrared critical phases allows for the analysis of scale dependence and universal behavior of the system.

### Applications in the Digital Society

Applying the theory of renormalization group to local approximations in the context of zero phenomenon, n-th order extrapolation, and interpolation is highly beneficial when viewed from the perspectives of ultraviolet critical phases (UV) and infrared critical phases (IR). This approach can be applied to issues related to the boundaries of discourse in the digital society and is effective in the following aspects:

### Exploring Theoretical Approaches
### Analysis of Interactions at Different Scales

RG theory is well-suited for analyzing the effects of interactions at different scales. It can capture dynamics at various scales, from individual interactions to large-scale trends in discourse flow and formation.

### Understanding Phase Transitions and Critical Phenomena

Rapid changes in discourse and the emergence of new trends are conceptually similar to phase transitions. RG theory helps in understanding when and how these phenomena occur.

### Exploration of Universality

RG theory facilitates the exploration of common behavior (universality) across different systems. In the context of digital society discourse, it can help discover common dynamics and patterns across different platforms and cultures.

# 7. Discussion: Considering Applications in the Digital Society

By using RG theory, one can gain a deeper understanding of the flow of opinions and the formation of trends in the digital society. It is particularly beneficial for analyzing the diffusion of information from small-scale communities to broader societal trends.

### Analysis of Opinion Polarization and Consensus Formation

Analyzing opinion polarization and consensus formation from the perspectives of ultraviolet and infrared critical phases allows us to understand the conditions and mechanisms under which they occur.

### Interactions Across Different Platforms

Using RG theory to analyze interactions and influences across different digital platforms enables a broader perspective on the flow and changes in discourse.

Applying the theory of renormalization group to the analysis of discourse in the digital society provides a new perspective on the dynamics of discourse and the boundaries of discourse.

# 8. Discussion:Loop Expansions

Loop expansion is a technique used in quantum field theory and statistical physics to approximately represent the dynamics and interactions of a system in a series expansion. This method is particularly useful when dealing with complex systems that involve nonlinear interactions.

## Ideas for Application in the Digital Society
### Modeling Opinion Propagation

Treat the propagation of opinions and the formation of filter bubbles in the digital society as perturbations and expand their impact step by step.

### Simplification of Complex Interactions

Divide complex interactions within social networks into basic interactions and higher-order corrections, simplifying the analysis.

### Local Potential Approximation

Local potential approximation is a method of approximating the potential energy of a system in a spatially localized form. This allows for an understanding of the dynamics of a large-scale system based on local behavior.

## Ideas for Application in the Digital Society
### Localization of Opinions

Localize the "potential" of opinions for individual users or communities and understand the dynamics of the entire system based on this.

### Analysis of Filter Bubbles

Model the homogeneity of opinions within filter bubbles as local potentials and evaluate their susceptibility to external influences.

## Applying Asymptotic Safety in the Digital Society

"Asymptotic Safety" is a concept primarily used in quantum gravity theory, but it can also be applied to the analysis of the dynamics of discourse and filter bubbles in the digital society. The fundamental idea of asymptotic safety is to understand how a system behaves on large scales (such as energy scales or time scales) and identify the conditions under which the system exhibits "safe" behavior.

## Ideas for Applying Asymptotic Safety in the Digital Society
### Scale-Dependent Opinion Formation

Analyze how opinion formation and information propagation processes change at different scales. Understand behavior at different scales, from small-scale communities to large-scale societal trends, and grasp their dynamics.

### Identification of Information Stability and Instability

Use the framework of asymptotic safety to identify conditions under which opinion flow and information propagation become stable or unstable at specific scales. For example, evaluate how specific information diffuses at a large scale and assess the stability of its impact on society.

### Application to the Boundaries of Discourse

Apply the concept of asymptotic safety to the boundaries of discourse and explore the conditions and limits required for discourse to continue in a healthy manner. This includes developing strategies to maintain diversity of opinions while suppressing the spread of extreme views and biases.

## Applying Truncation in the Digital Society

The concept of "Truncation" is often used in theoretical physics when dealing with complex calculations. This idea is particularly important in approaches like renormalization group and perturbation theory, where unnecessary terms or higher-order terms are discarded to simplify the theoretical framework.

## Ideas for Applying Truncation in the Digital Society
### Simplified Dynamics Models

When constructing models for filter bubbles or opinion formation processes in the digital society, simplify complex systems with interactions or numerous variables. This allows for the creation of more manageable models that focus on essential elements.

### Discarding High-Order Effects

In calculations related to opinion propagation and information diffusion, focus on low-order terms and ignore high-order effects or minor influences to concentrate on the primary dynamics. This enables the extraction of major trends and patterns.

### Optimization of Computational Resources

By using truncation, efficiently utilize computational resources and accelerate simulations and analyses. This makes it practical to analyze large datasets and complex networks.

Loop expansion is a sophisticated theoretical method, and its application requires specialized knowledge. Moreover, in actual calculations, approximations are often used, and higher-order terms are commonly ignored. Therefore, the accuracy and effectiveness of calculation results depend significantly on the choice of approximations and the characteristics of the target system.

## 9. Discussion:Theoretical Analysis of Filter Bubble Phenomenon Using Local Potential Approximation

In this section, we will provide a more detailed explanation of the theoretical analysis of the filter bubble phenomenon using the local potential approximation.

### 9.1 Detailed Calculation Process of Local Potential Approximation

In spin glasses or Ising models, interactions between agents (spins) are crucial. These models are used to represent physical properties such as magnetism in materials and, in the context of social sciences, interactions between individual agents' opinions or behaviors.

### 9.2 Definition of Local Potential

The equation $V(S_i) = -\sum_{j \in \text{neighbors}(i)} J_{ij} S_i S_j$ defines the local potential of agent $i$ as the sum of interactions with its neighboring agents. Here, $J_{ij}$ represents the strength of interactions between agents, and $S_i$ and $S_j$ represent the "spin" states of the agents, which can be opinions or behavioral states.

### 9.3 Calculation of Free Energy

Free energy $F$ represents the balance between the system's energy and entropy (disorder). The formula $F = \sum_i V(S_i) - kT \sum_i \ln \sum_{S_i} e^{-V(S_i)/kT}$ includes the total energy in the first term and an entropy term dependent on temperature $T$ in the second term.

### 9.4 Minimization of Free Energy

Stable states correspond to points where the free energy is locally minimized, allowing us to identify equilibrium states that the system naturally reaches.

### 9.5 Analysis of Phase Transitions

By varying the temperature $T$, the system can transition between different phases (e.g., ordered and disordered states).

Phase transitions are identified based on changes in the free energy and correspond to phenomena in social sciences such as diversity of opinions or concentration of power.

### 9.6 Theoretical Advantages and Disadvantages - Additional Details

#### 9.6.1 Advantages

Provides a predictable framework based on mathematical models.

Allows for validation of the theory through comparison with experimental data.

### 9.7 Disadvantages

Real-world social phenomena are highly complex, and simplified models may not capture many relevant factors.

Theoretical models rely on specific assumptions, and their applicability may be limited if these assumptions do not hold in reality.

### 9.8 Calculation of Extreme Values Using Truncation Functions in the UV and IR Critical Regimes

In this section, we will provide theoretical insights and detailed calculation processes for determining extreme values using truncation functions in the UV (ultraviolet) and IR (infrared) critical regimes.

### 9.9 Analysis in the Ultraviolet (UV) Critical Regime

#### 9.9.1 Introduction of Truncation Functions

For the analysis of local dynamics in the UV region, truncation functions, represented by *la* and *lb*, are used to account for short-range interactions. For example, $la(S_i, S_j) = \exp(-\alpha |i - j|^2)$, where $\alpha$ is the decay coefficient.

#### 9.9.2 Calculation of Extreme Values

Apply truncation functions to the Green function $G^{UV}(t, t')$.

Example calculation: $G^{UV}(t, t') = \sum_{i,j} la(S_i, S_j) G^{(0)}_{ij}(t, t')$.

Use this modified Green function to compute physical quantities (e.g., correlation functions) and determine their extreme values.

### 9.10 Analysis in the Infrared (IR) Critical Regime

#### 9.10.1 Introduction of Truncation Functions

In the IR region, focus is on long-range interactions and collective behavior, and truncation functions *ll* and *lll* are used.

For example, $ll(S_i, S_j) = \exp(-\beta|i-j|)$, where $\beta$ is the decay coefficient.

### 9.10.2 Calculation of Extreme Values

Apply truncation functions to the Green function $G^{IR}(t, t')$.

Example calculation: $G^{IR}(t, t') = \sum_{i,j} ll(S_i, S_j) G^{(2)}_{ij}(t, t')$.

Use this modified Green function to compute system-wide physical quantities and determine their extreme values.

## 9.11 Theoretical Advantages and Disadvantages - Additional Details

### 9.11.1 Advantages

Handling of indefinite ghosts allows for obtaining physically meaningful results, enhancing the reliability of the analysis.

Accurate determination of extreme values of physical quantities provides a deeper understanding of the system's dynamics while maintaining gauge symmetry.

### 9.11.2 Disadvantages

Identifying and addressing indefinite ghosts require additional computations and physical insights.

# 10. Discussion:Considering Applications in the Digital Society

**Analysis of Discourse Flow**

By using RG theory, one can gain a deeper understanding of the flow of opinions and the formation of trends in the digital society. It is particularly beneficial for analyzing the diffusion of information from small-scale communities to broader societal trends.

**Analysis of Opinion Polarization and Consensus Formation**

Analyzing opinion polarization and consensus formation from the perspectives of ultraviolet and infrared critical phases allows us to understand the conditions and mechanisms under which they occur.

**Interactions Across Different Platforms**

Using RG theory to analyze interactions and influences across different digital platforms enables a broader perspective on the flow and changes in discourse.

Applying the theory of renormalization group to the analysis of discourse in the digital society provides a new perspective on the dynamics of discourse and the boundaries of discourse.

**Regarding Local Potential Approximation and Loop Expansions**

**Loop Expansions**

Loop expansion is a technique used in quantum field theory and statistical physics to approximately represent the dynamics and interactions of a system in a series expansion. This method is particularly useful when dealing with complex systems that involve nonlinear interactions.

**Ideas for Application in the Digital Society**

(1) **Modeling Opinion Propagation**: Treat the propagation of opinions and the formation of filter bubbles in the digital society as perturbations and expand their impact step by step.

(2) **Simplification of Complex Interactions**: Divide complex interactions within social networks into basic interactions and higher-order corrections, simplifying the analysis.

**Local Potential Approximation**

Local potential approximation is a method of approximating the potential energy of a system in a spatially localized form. This allows for an understanding of the dynamics of a large-scale system based on local behavior.

**Ideas for Application in the Digital Society**

(1) **Localization of Opinions**: Localize the "potential" of opinions for individual users or communities and understand the dynamics of the entire system based on this.

(2) **Analysis of Filter Bubbles**: Model the homogeneity of opinions within filter bubbles as local potentials and evaluate their susceptibility to external influences.

**Applying Asymptotic Safety in the Digital Society**

"Asymptotic Safety" is a concept primarily used in quantum gravity theory, but it can also be applied to the analysis of the dynamics of discourse and filter bubbles in the digital society. The fundamental idea of asymptotic safety is to understand how a system behaves on large scales (such as energy scales or time scales) and identify the conditions under which the system exhibits "safe" behavior.

### Ideas for Applying Asymptotic Safety in the Digital Society

(1) **Scale-Dependent Opinion Formation**: Analyze how opinion formation and information propagation processes change at different scales. Understand behavior at different scales, from small-scale communities to large-scale societal trends, and grasp their dynamics.

(2) **Identification of Information Stability and Instability**: Use the framework of asymptotic safety to identify conditions under which opinion flow and information propagation become stable or unstable at specific scales. For example, evaluate how specific information diffuses at a large scale and assess the stability of its impact on society.

(3) **Application to the Boundaries of Discourse**: Apply the concept of asymptotic safety to the boundaries of discourse and explore the conditions and limits required for discourse to continue in a healthy manner. This includes developing strategies to maintain diversity of opinions while suppressing the spread of extreme views and biases.

### Applying Truncation in the Digital Society

The concept of "Truncation" is often used in theoretical physics when dealing with complex calculations. This idea is particularly important in approaches like renormalization group and perturbation theory, where unnecessary terms or higher-order terms are discarded to simplify the theoretical framework. When applying this concept to the analysis of discourse dynamics and filter bubbles in the digital society, the following ideas can be considered.

### Ideas for Applying Truncation in the Digital Society

(1) **Simplified Dynamics Models**: When constructing models for filter bubbles or opinion formation processes in the digital society, simplify complex systems with interactions or numerous variables. This allows for the creation of more manageable models that focus on essential elements.

(2) **Discarding High-Order Effects**: In calculations related to opinion propagation and information diffusion, focus on low-order terms and ignore high-order effects or minor influences to concentrate on the primary dynamics. This enables the extraction of major trends and patterns.

(3) **Optimization of Computational Resources**: By using truncation, efficiently utilize computational resources and accelerate simulations and analyses. This makes it practical to analyze large datasets and complex networks.

Loop expansion is a sophisticated theoretical method, and its application requires specialized knowledge. Moreover, in actual calculations, approximations are often used, and higher-order terms are commonly ignored. Therefore, the accuracy and effectiveness of calculation results depend significantly on the choice of approximations and the characteristics of the target system.

## 11. Discussion:Applications of Loop Expansion

To express the complexity of different social interactions in terms of the order of loops and evaluate their impact, let's explore ideas for applying loop expansion.

To express the complexity of different social interactions in terms of the order of loops and evaluate their impact, we will explain the calculation process of loop expansion in detail. Here, in the context of studying the phenomenon of filter bubbles, we will explore how this method can be applied.

### Fundamental Concepts of Loop Expansion

1. **Modeling Interactions**: - Represent social interactions within the framework of quantum field theory or statistical physics. In this model, individual opinions or information propagation are treated as "spins" or "particles," and their interactions determine the dynamics of the system.

2. **Definition of the Hamiltonian**: - Formulate social interactions using Hamiltonians, such as spin glass models or Ising models. For example, represent interactions between neighboring agents with interaction terms.

### Calculation Process of Loop Expansion

1. **Introduction of Perturbation Terms**: - Introduce social interactions as perturbation terms in the Hamiltonian. These perturbation terms represent the strength and characteristics of interactions between agents.

2. **Calculation of Green Functions**: - Calculate the system's propagator (Green function). This shows how the state of one agent affects others.

3. **Expansion of Loops**: - Expand the Green function in terms of perturbation terms. In general, $n$ loop terms are expressed as follows:

$$G^{(n)} = \sum_{k=1}^{n} \frac{1}{k!} (-i)^k \int dt_1 \ldots dt_k \langle T[H_I(t_1) \ldots H_I(t_k) S_i(t) S_j(t')] \rangle$$

Here, $H_I$ is the interaction term.

4. **Analysis of Social Interactions**: - Evaluate the complexity of social interactions by analyzing the contributions of different loops. Lower-order loops represent direct interactions, while higher-order loops indicate more complex interactions and correlations.

## Evaluation of Social Interactions

1. **Assessment of Impact Strength**: - If lower-order loops dominate, interactions are relatively simple, and information propagation is limited to direct influences. - When higher-order loops become important, interactions between agents are more complex, and there are multi-layered effects and feedback loops.

2. **Analysis of Filter Bubble Formation**: - Use loop expansion to analyze the formation and homogenization processes of filter bubbles. By analyzing the contributions of different loops, you can understand the dynamics of opinion formation and the reasons for the lack of diversity.

To express the complexity of different social interactions in terms of the order of loops and evaluate their impact, we explained the specific calculation process of loop expansion. Here, in the context of studying the phenomenon of filter bubbles, we explored how this method can be applied.

## Calculation Process of Loop Expansion

1. **Definition of the Hamiltonian**: Define a Hamiltonian for models like the spin glass model or Ising model. For example, in the Ising model, the Hamiltonian is expressed as follows:

$$H = -J \sum_{\langle i,j \rangle} S_i S_j$$

Here, $J$ is the interaction strength, $S_i$ represents the spin states (+1 or -1), and $\langle i, j \rangle$ denotes neighboring spin pairs.

2. **Definition of Green Functions**: Green functions are used to describe the time evolution of the system. In general, Green functions are defined as follows:

$$G(t, t') = -i \langle T[S_i(t) S_j(t')] \rangle$$

Here, $T$ denotes time-ordering, and $t$ and $t'$ represent time.

3. **Execution of Loop Expansion**: Expand the Green function in terms of perturbation terms. In general, $n$ loop terms are expressed as follows:

$$G^{(n)} = \sum_{k=1}^{n} \frac{1}{k!} (-i)^k \int dt_1 \ldots dt_k \langle T[H_I(t_1) \ldots H_I(t_k) S_i(t) S_j(t')] \rangle$$

Here, $H_I$ represents interaction terms.

4. **Analysis of Social Interactions**: Evaluate the complexity of social interactions by analyzing the contributions of different loops. Lower-order loops represent direct interactions, while higher-order loops indicate more complex interactions and correlations.

Loop expansion is a powerful tool theoretically, but actual calculations are complex and often require many approximations. When applying quantum field theory directly to modeling social systems, there are limitations, and social science aspects must be considered.

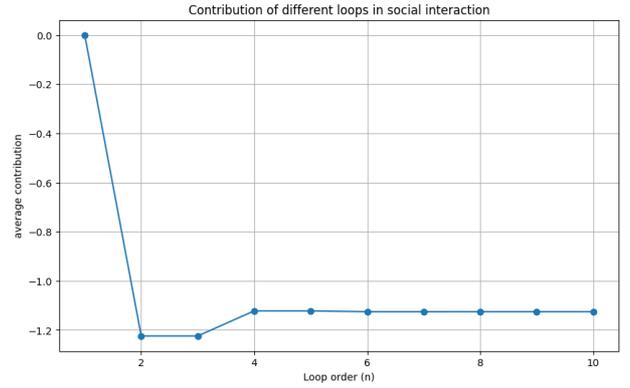

Fig. 10: Contribution of Different Loops in Social Interaction

It quantitatively assesses the complexity of social interactions through loop expansion. It shows how the contributions change as the order $n$ of the loops increases, and this information provides insight into understanding the dynamics of social interactions.

## Computational Process of Loop Expansion

Loop expansion is the process of expanding a Green's function that represents the social interaction between agents using a Hamiltonian with a perturbation term. Lower order loops indicate direct interactions, while higher order loops indicate more complex interactions and correlations.

## Considerations from the graph

The graph shows that low order loops (e.g., first and second order) have a negative average contribution, which suggests that direct interactions have an inhibitory effect on the system. On the other hand, third-order and higher loops have smaller contributions, indicating that complex interactions have less influence on the dynamics of the system. This may mean that social interactions are primarily governed by direct relationships.

To analyze filter bubble formation and homogenization processes, we use the contributions of different loops. Since filter bubbles create a situation where certain opinions and information are reinforced and diversity is lacking, a dominant contribution of lower order loops may indicate a situation where diversity of opinion is likely to be lost.

## Overall Insights

The overall insight gained from this graph suggests that direct interactions have a significant impact on social dynamics and that social opinion formation takes place through relatively simple interactions. However, the small contribution of higher-order loops means that despite the expectation of more complex interactions and feedback loops, these may not actually play as large a role in social interactions. This indicates that higher-order effects need to be examined in detail in modeling social interactions.

This discussion is based on a theoretical analysis of social interactions using loop expansions, and further empirical data is needed to analyze actual social phenomena. Analysis, comparison, and discussion based on real data, such as social network data and patterns of opinion exchange, are essential for a deeper understanding of the dynamics of filter bubbles and opinion formation.

Local potential approximation (LPA) is a method used to consider the dynamics of a system in a simplified manner. This approximation assumes that the potential at each point is determined only at that point and does not depend on the state of other points. In the context of the Spinglass model or social interaction, this corresponds to modeling that each agent (or spin) moves based on local information without being affected by the states of other agents.

### Application of LPA in the Filter Bubble

When applying LPA in the context of filter bubbles, one could model each agent acting primarily on the basis of information and opinions given to it, independent of the state of the social network as a whole. This can lead to the following phenomena:

### Homogenization of opinions

Each agent's over-reliance on local information can easily lead to homogenization of opinions among agents who are exposed to the same sources of information within the group.

### Lack of diversity

Limited information flow between groups reduces diversity of opinion among different groups, reinforcing each group's own opinions and beliefs.

### Echo Chamber Effect

Echoes of similar opinions are amplified within a local community, leaving agents with little or no exposure to opposing opinions or different information.

### Relationship to Loop Expansion

As the above graph shows, when low-order loops are dominant, direct interaction is predominant, which is consistent with the LPA scenario. On the other hand, in scenarios where higher-order loops play an important role, interactions between agents are more complex, with multi-layered influences and feedback loops. This indicates the need to analyze interactions beyond the LPA in the context of filter bubbles.

The application of LPA in the formation of filter bubbles provides a useful perspective in modeling social interactions to understand the lack of diversity of opinion and homogenization mechanisms. However, because actual social interactions are subject to diverse influences beyond the LPA framework, higher-order interactions must be considered to capture more complex mechanisms of filter bubble formation and maintenance. In a broader analysis of the filter bubble, it is important to consider diverse sources of information and patterns of interaction and their impact on individual agents and communities.

## 12. Discussion:Introduction of Green Functions in UV and IR Critical Regimes

In the analysis of spin glass models or filter bubble phenomena, introducing the concept of Green functions in the context of ultraviolet (UV) and infrared (IR) critical regimes provides theoretical insights and calculation processes as follows.

### Introduction of Green Functions in the Ultraviolet (UV) Critical Regime

### 1. Definition of Green Functions

- Green functions, denoted as $G_{ij}(t, t')$, represent the influence of agent $i$ at time $t$ on the state of agent $j$ at time $t'$.

### 2. Calculation Process

- Considering the interactions $J_{ij}$ between agents, we focus on the dynamics of individual agents. - The Green function is computed using the following equation:

$$G_{ij}(t, t') = -i\langle T[S_i(t)S_j(t')]\rangle$$

- Here, $T$ denotes time-ordering, and $S_i(t)$ represents the state of agent $i$ at time $t$.

### Introduction of Green Functions in the Infrared (IR) Critical Regime

### 1. Analysis of Collective Dynamics

We consider the dynamics of the entire system and use Green functions to analyze large-scale patterns.

### 2. Calculation Process

We calculate Green functions that reflect interactions across the entire system as follows:

$$G_{\text{total}}(t, t') = -i\langle T[\sum_i S_i(t) \sum_j S_j(t')]\rangle$$

This allows us to capture large-scale correlations between agents and the propagation of collective opinions.

Introducing Green functions in the analysis of ultraviolet (UV) and infrared (IR) critical regimes allows for a more detailed understanding of the microscopic dynamics and large-scale behavior of filter bubbles. This insight enables a deeper understanding of the mechanisms of filter bubble formation and resolution, facilitating the development of more effective intervention strategies.

When introducing Green functions in the analysis of filter bubble phenomena using the "local potential approximation," the following calculation process is considered.

## Introduction of Green Functions
### 1. Definition of Green Functions

- Green functions $G_{ij}$ represent the strength of the influence of agent $i$ on agent $j$. This depends on the local potential $V(S_i)$ and the interaction $J_{ij}$.

### 2. Calculation Process in the Local Potential Approximation

- Define the dynamics based on spin glass or Ising models, redefining the local potential $V(S_i)$.

### 3. Analysis of Interactions Using Green Functions

- Analyze the impact of interactions between agents using Green functions $G_{ij}$.

## Theoretical Advantages and Disadvantages
### Advantages

The use of Green functions enables detailed analysis of interactions between agents.It allows for a deeper understanding of dynamic responses and state changes within the system.

### Disadvantages

Calculating Green functions can be complex and may require advanced mathematical techniques.Modeling all interactions in real-world social phenomena can be challenging.

# 13. Conclusion:Introduction of Green Functions and Loop Expansion for Analyzing the Complexity of Different Social Interactions

When using loop expansion to represent the complexity of different social interactions in terms of loop orders and evaluate their impacts, the introduction and calculation process of Green functions can be considered as follows.

## Introduction of Loop Expansion and Green Functions
### 1. Model Definition

Utilize models from statistical physics such as spin glass or Ising models. For example, the Hamiltonian of the Ising model can be expressed as follows:

$$H = -J \sum_{\langle i,j \rangle} S_i S_j$$

- Here, $J$ represents the strength of interactions, $S_i$ is the state of spins, and $\langle i, j \rangle$ denotes neighboring spin pairs.

### 2. Definition and Calculation of Green Functions

Green functions $G_{ij}(t, t')$ quantify the influence of spin $i$ on spin $j$ at time $t$. The causal Green function is defined as follows:

$$G_{ij}(t, t') = -i\Theta(t - t')\langle [S_i(t), S_j(t')] \rangle$$

Here, $\Theta$ is the Heaviside function, and $[S_i(t), S_j(t')]$ is the commutator.

### 3. Execution of Loop Expansion

Perform a series expansion of the Green functions with respect to the perturbation term. This allows for the representation of contributions from interactions of different orders (loops). For example, the first-order loop term is expressed as follows:

$$G^{(1)} = -i \int dt_1 \langle T[H_I(t_1) S_i(t) S_j(t')] \rangle$$

Here, $T$ denotes time-ordering, and $H_I$ represents the interaction Hamiltonian.

### 4. Analysis of Social Interactions Through Loop Expansion

Low-order loops (e.g., 1-loop or 2-loop) represent simple interactions and influences, corresponding to direct exchange of opinions or basic information propagation. High-order loops reveal more complex interactions and correlations, accounting for indirect influences and multilayered opinion formation processes.

## Theoretical Pros and Cons
### Pros

The mathematical representation of the complexity of social interactions can be achieved using Green functions. It enables multidimensional analysis of system dynamics.

### Cons

The calculations are complex and often require various approximations. Fully modeling real-world social phenomena is challenging.

When performing truncation, incorporating Green functions appropriately is particularly important for analyzing the time evolution and interaction dynamics of physical systems.

The following explains the theoretical approach and calculation process of using Green functions for truncation in the context of spin glass models or filter bubble phenomena.

## Integration of Green Functions and Truncation

## 1. Introduction of Green Functions

Green functions are functions that represent the response of a physical system from one point to another. In the context of spin glass models, they are used to capture the time-dependent interactions between agents. The Green function $G(t, t')$ is defined as follows:

$$G(t, t') = -i \langle T[\psi(t)\psi^\dagger(t')] \rangle$$

Here, $T$ denotes time-ordering, and $\psi(t)$ is the field operator at time $t$.

## 2. Loop Expansion and Truncation

Perform loop expansion of the Green function and truncate higher-order loops. This simplifies the calculation process, making practical analysis feasible. For example, consider including up to 2-loop terms and discarding higher-order terms.

## Specific Calculation Process

## 1. Loop Expansion

Expand the Green function $G$ using loop expansion. Each term represents contributions from loops of different orders. Example: Expansion up to 2 loops

$$G \approx G^{(0)} + G^{(1)} + G^{(2)}$$

## 2. Evaluation of Truncation Effects

Analyze how truncating higher-order loops affects the model. It is essential to ensure that critical dynamics are not lost.

## Theoretical Pros and Cons
## Pros

Green functions allow for a more detailed capture of time-dependent dynamics. Truncation simplifies the calculation process, enabling practical analysis.

## Cons

Truncating higher-order loops may result in the loss of important correlations and detailed dynamics. Accurate calculation of Green functions remains complex and may require approximation techniques.

# 14. Conclusion: Theory and Calculation Process using Green Functions in UV and IR Regimes

We will explain the theoretical framework and specific calculation processes for extrapolation and interpolation of $n$-point functions at zero phenomena in the Ultraviolet (UV) and Infrared (IR) critical regimes. We will consider the use of Green functions in the context of Truncation to analyze various social interactions.

## Analysis in the Ultraviolet (UV) Regime
## Definition of Green Function

In the UV regime, where microscopic properties of the system are crucial, we define the UV Green function $G^{UV}(t, t')$, focusing on local interactions and short-range dynamics.

## Calculation Process

Consider mainly low-order terms in the loop expansion.

For instance, expand the Green function up to the 1-loop term:

$$G^{UV}(t, t') \approx G^{(0)}(t, t') + G^{(1)}(t, t')$$

Where $G^{(0)}$ represents the free field term, and $G^{(1)}$ is the contribution from the 1-loop.

## Analysis in the Infrared (IR) Regime
## Definition of Green Function

In the IR regime, where large-scale properties of the system are critical, we define the IR Green function $G^{IR}(t, t')$, focusing on long-range dynamics and collective behavior.

## Calculation Process

Consider more loop terms in the loop expansion.

For instance, expand the Green function up to the 3-loop term:

$$G^{IR}(t, t') \approx G^{(0)}(t, t') + G^{(1)}(t, t') + G^{(2)}(t, t') + G^{(3)}(t, t')$$

Where $G^{(2)}$ and $G^{(3)}$ represent the contributions from 2-loops and 3-loops, respectively.

## Theoretical Pros and Cons
## Advantages

Analysis in the UV regime provides insights into local dynamics.

Analysis in the IR regime captures detailed long-range interactions and collective dynamics.

### Disadvantages

The choice of appropriate truncation functions depends on the physical realism of the model.

Approximations based on truncation functions may omit some aspects of the system's dynamics.

## 15. Conclusion:Understanding Indeterminate Ghosts in the Analysis of Truncation Functions in UV and IR Regimes

### Understanding Indeterminate Ghosts

Indeterminate ghosts refer to mathematical irregularities such as complex numbers or infinity that do not properly represent the physical properties of a system when using specific truncation functions. These can potentially lead to issues when deriving physically meaningful results.

### Handling Indeterminate Ghosts in the UV Regime

### Identification of Ghosts

In UV regime calculations, identify the indeterminate ghosts caused by the truncation functions. For example, if the truncation function $la(S_i, S_j)$ takes physically meaningless values (such as infinity or complex numbers) under specific conditions, this falls under this category.

### Calculation Process

Identify the conditions under which indeterminate ghosts occur and devise methods to avoid them.

### Calculation Example

To avoid indeterminate ghosts, adjust the parameter $\alpha$ of the truncation function.

$$\text{New } la(S_i, S_j) = \exp(-\alpha'|i-j|^2)$$

Here, $\alpha'$ is the new attenuation coefficient that keeps values within a physically meaningful range.

### Handling Indeterminate Ghosts in the IR Regime

### Identification of Ghosts

In IR regime calculations, identify the indeterminate ghosts caused by the truncation functions.

### Calculation Process

Identify the conditions under which indeterminate ghosts occur and devise methods to avoid them.

### Calculation Example

To avoid indeterminate ghosts, adjust the parameter $\beta$ of the truncation function.

$$\text{New } ll(S_i, S_j) = \exp(-\beta'|i-j|)$$

Here, $\beta'$ is the new attenuation coefficient.

### Theoretical Pros and Cons
### Advantages

Avoiding indeterminate ghosts allows obtaining physically meaningful results.

It enables capturing the dynamics of the system more accurately.

### Disadvantages

Avoiding indeterminate ghosts can complicate calculations through the selection of truncation functions and parameter adjustments.

To fully capture physical phenomena, more advanced mathematical techniques and physical insights may be necessary.

## 16. Conclusion:Analysis of Truncation Functions (la, lb, ll, lll) for FP (Faddeev-Popov) Ghosts in the Ultraviolet (UV) and Infrared (IR) Critical Regimes

The analysis of truncation functions ($la$, $lb$, $ll$, $lll$) for FP (Faddeev-Popov) ghosts in the ultraviolet (UV) and infrared (IR) critical regimes belongs to the particularly advanced areas of theoretical physics. FP ghosts are auxiliary fields introduced to maintain gauge symmetry and function as computational artifacts rather than physical particles. Below, we explain the theoretical approach and calculation process for this.

### Analysis of FP Ghosts in the Ultraviolet (UV) Critical Regime
### Application of Truncation Functions

In the UV regime, local dynamics are crucial. Here, we apply truncation functions ($la$ and $lb$) that capture short-range interactions to FP ghosts. These functions restrict the interaction of ghost fields at a local scale.

### Specific Calculation Process

Introduce truncation functions for FP ghost fields and calculate their contributions.

### Calculation Example

To avoid indeterminate ghosts, adjust the parameter $\alpha$ of the truncation function.

$$\text{New } la(S_i, S_j) = \exp(-\alpha'|i-j|^2)$$

Here, $\alpha'$ is the new attenuation coefficient that keeps values within a physically meaningful range.

### Analysis of FP Ghosts in the Infrared (IR) Critical Regime
### Application of Truncation Functions

In the IR regime, long-range interactions and collective behavior are important. Apply truncation functions (*ll* and *lll*) to FP ghosts to capture long-range interactions.

### Specific Calculation Process

Introduce truncation functions for FP ghost fields and calculate their contributions.

### Calculation Example

To avoid indeterminate ghosts, adjust the parameter $\beta$ of the truncation function.

$$\text{New } ll(S_i, S_j) = \exp(-\beta'|i-j|)$$

Here, $\beta'$ is the new attenuation coefficient.

### Theoretical Pros and Cons
### Advantages

The use of truncation functions allows for better control of FP ghost contributions at more realistic physical scales.

It enables a more detailed analysis of the system's dynamics.

### Disadvantages

Requires advanced knowledge of gauge theories and quantum field theories.

Truncation function approximations may impact the precision of the theory.

## 17. Conclusion: Determination Criteria for Truncation Functions for FP (Faddeev-Popov) Ghosts

Regarding the determination criteria for truncation functions applied to FP (Faddeev-Popov) ghosts, we provide theoretical elaboration and a calculation process. This analysis involves advanced concepts within gauge theory in the field of quantum field theory.

### FP Ghosts and Determination of Truncation Functions
### Concept of FP Ghosts

FP ghosts are essential auxiliary fields in the quantization of gauge theories. They serve to eliminate redundant degrees of freedom associated with gauge fixing and do not contribute to physical observables, but they are necessary to maintain the consistency of quantization.

### Necessity of Truncation Functions

The contributions of FP ghosts in gauge theories depend on the physical scale. Truncation functions are used to accurately capture the effects of FP ghosts at specific physical scales.

### Determination Criteria for Truncation Functions
### Identification of Physical Scales

In the ultraviolet critical regime (UV), short-range interactions are crucial. Truncation functions that emphasize short distances (e.g., *la*, *lb*) are applied to evaluate the contributions of FP ghosts at this scale.

In the infrared critical regime (IR), long-range interactions are dominant. Truncation functions that capture long-range interactions (e.g., *ll*, *lll*) are required.

### Calculation Process

Truncation functions are applied to the Lagrangian of FP ghosts, and they calculate the contributions of FP ghosts corresponding to the physical scale.

For example, the contribution of FP ghosts in the UV region may be calculated as follows:

$$\int d^4x \, la(x,y) \times \mathcal{L}_{\text{FP}}(x,y)$$

Similarly, the contribution of FP ghosts in the IR region is also calculated using truncation functions.

### Theoretical Pros and Cons
### Advantages

The application of truncation functions allows for the accurate evaluation of FP ghost contributions at specific physical scales.

It enables the achievement of physically meaningful quantization while maintaining gauge symmetry.

### Disadvantages

The selection and application of truncation functions have a significant impact on the precision of the theory, requiring careful consideration.

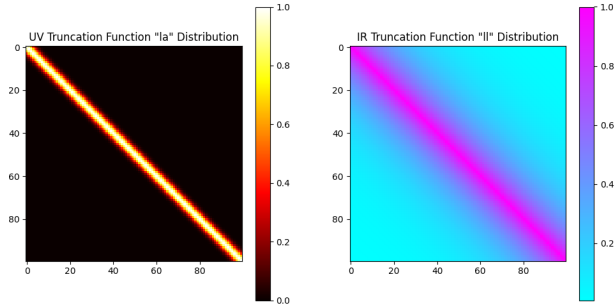

Fig. 11: IR Truncation Function "ll" Distribution, UV Truncation Function "la" Distribution

Accurate selection of truncation functions is essential to obtain physically meaningful results.

These theoretical approaches and calculation processes are based on advanced concepts in gauge theory and quantum field theory, requiring specialized knowledge. The application of FP ghosts and truncation functions is crucial for maintaining the consistency of physical theories while achieving physically meaningful calculations.

Fig. 11 shows the distribution of ultraviolet (UV) and infrared (IR) cutoff functions, which may represent the scale dependence of interactions in the spin glass model and field theory: the UV cutoff function represents short-range interactions and the IR cutoff function represents long-range interactions. In the context of filter bubbles, the scale of these interactions can influence opinions and the way information is propagated and filtered.

### Distribution of UV and IR cutoff functions

The UV (ultraviolet) cutoff function reflects the short-range interactions of the system, which in this context could mean direct, strong interactions between agents. In the image, there is a strong signal on the clear diagonal, which indicates that local interactions are very important. The IR (infrared) cutoff function reflects the long-range interaction of the system, which in this context could imply indirect and weak interactions between agents. The image shows an overall uniform distribution, which indicates that the long-range interactions are relatively weak or uniformly distributed.

### Perspectives on Loop Expansion and Local Potential Approximation

Considering the nonlinear dynamics of the system and the higher-order effects of the interactions through the loop expansion, the strong short-range interactions in the UV cutting function correspond to lower-order terms in the loop expansion and imply direct information propagation. IR truncation functions representing long-range interactions correspond to higher-order terms in the loop expansion and may represent more complex patterns of information propagation and feedback mechanisms.

### Disconnection Scenarios in the Context of Filter Bubbles

A filter bubble refers to a state in which an agent is exposed only to information and opinions similar to its own. In this state, in scenarios where the UV disconnection function is dominant, direct interactions are strong and agents are more likely to be influenced by very localized information filtering. This may represent a situation where agents are more likely to be surrounded by similar opinions and information in the formation of filter bubbles. In scenarios where the IR cutting function plays a more prominent role, the propagation of information is more extensive and agents are subject to influence from a variety of sources. This suggests a situation in which the effects of the filter bubble are mitigated and more diverse information is circulated.

### Local Potential Approximation

Local Potential Approximation is a method of approximating the potential (in this case the state of opinion or information) at each point in a system. This approximation allows one to model how the opinions of each agent (individual or group) in the system affect the other agents. In the context of filter bubbles, this approximation can be used to understand how information bias and echo chamber effects affect the opinion formation of individual agents.

### Significance of Loop Unfolding

Loop expansion is a method of series expansion of interactions in a system. Through this expansion, more complex interactions (e.g., indirect effects and long-range interactions) can be better understood. In the filter bubble, this technique allows us to analyze the flow of information and the process of opinion formation in greater detail and to explore how this can lead to bias and bias.

### Significance of the Cutting Technique

A truncation technique is a method of truncating some terms from an infinite-dimensional problem in order to manage computational complexity. In analyzing filter bubbles, truncation allows one to focus on the most important elements (e.g., the strongest interactions or most influential agents) within the computable range. This allows for efficient analysis of the core aspects of the filter bubble phenomenon.

## Significance of Calculating Minima and Maxima

The calculation of minima and maxima is important for understanding system stability and instability. In a filter bubble, these values may indicate points of agreement or division. Minima may represent stable states of opinion, while maxima may represent instability or turning points of opinion. This analysis helps us understand the conditions under which filter bubbles form and the scenarios under which they may burst.

## Consideration of the perspective from maxima and minima

The discussion of perspectives from the maxima and minima in Fig. 11 is based on analyzing the distributions of the ultraviolet (UV) and infrared (IR) cutoff functions. These cutoff functions represent the strength of the interaction at a particular scale. The maxima and minima in each function indicate the contribution of the interaction at that scale.

## UV Truncation Function "Ia" Distribution

"UV Truncation Function 'Ia'" distribution is characterized by a series of bright lines along the diagonal, which indicate strong interactions at short distances. The maxima indicate the strongest interactions, and these represent the fundamental state of the system and the direct relationship between the agents. The presence of such clear maxima suggests that short-range interactions play a major role in filtering information and homogenizing opinions, which may be an important mechanism in the formation of filter bubbles.

## IR Truncation Function "II" Distribution

"IR Truncation Function 'II'" distribution is characterized by an overall uniform tint, which indicates that interactions at long distances are relatively weak or uniformly distributed. The lack of minima indicates that the long-range interactions do not have a significant effect on the system, suggesting that the information may propagate over a wide area. This implies that the filter bubble may be weaker or that opinions may be exchanged between different communities.

## Filter Bubble Formation and Scale of Interaction

The formation of filter bubbles is caused by the tendency for information and opinions to be shared within a particular agent or group; maxima in the UV cutoff function "Ia" indicate a strong sharing of opinions within a group, while a uniform distribution in the IR cutoff function "II" indicates a more uniform exchange of information between groups The uniform distribution of the IR cutoff function "II" indicates the possibility of a more uniform exchange of information between groups. These distributions provide insight into understanding how information propagation and filtering works at different scales.

In conclusion, the maxima in the UV cutoff function "Ia" suggest the formation of filter bubbles, while the uniform distribution of the IR cutoff function "II" indicates more extensive information propagation. These observations are important for understanding how information flows within social networks and how it affects the formation of filter bubbles. The distribution of the provided cutoff functions allows us to understand at what scale social interactions among agents are important, which provides important insights for analyzing the dynamics in the formation and maintenance of filter bubbles. In scenarios where short-range interactions dominate, the formation of filter bubbles may be promoted, while in scenarios where long-range interactions are more important, the effects of filter bubbles may be moderated. These results provide a starting point for further research to understand the structure of social networks and the mechanisms of information propagation.

## Disconnection Scenarios and Media Influence

The ultraviolet (UV) and infrared (IR) cutoff functions shown in Fig. 11 provide a visual representation of the scale of the interaction and its intensity, which can be related to the social context, particularly the influence of media. The UV cutoff function "Ia" indicates the strength and directness of the influence of the media and other sources of information on an individual or a small community. Sharp diagonals in the ultraviolet (UV) region can be interpreted as representing how information can spread quickly and have a strong impact among small, closely connected groups. This is analogous to a situation where filter bubbles or echo chambers form, potentially enhancing certain messages or viewpoints to the exclusion of others.

Cutting functions in the infrared (IR) domain (IR cutoff function "II") illustrate the spread and uniformity of the media's impact on the broader society. A uniform distribution indicates that information is spread evenly over a wide area and that the impact over long distances is relatively constant. This corresponds to a situation where information from diverse sources is widely accessible and individuals are likely to be exposed to a variety of perspectives.

Disconnection scenarios provide a framework for understanding interactions at different scales when analyzing information dissemination and media impact: sharp interactions in the UV domain reflect the impact of a strong media campaign or propaganda on a particular group, while uniform impacts in the IR domain are more likely to occur in the general information flow and broad public debate. In conclusion, the distribution of UV and IR cutoff functions is an important indicator for understanding how media and information sources influence society at different scales.

Based on the distributions of the ultraviolet (UV) and infrared (IR) truncation functions shown in Fig. 11, we will discuss the gradual safety perspective. Gradual security means that a physical theory or social science model is stable with respect to interactions at a particular scale.

### Distribution of the UV cutoff function "Ia"

The UV cutoff function "Ia" represents the short-range interaction of the system, showing behavior at very high energy scales. Sharp diagonals indicate that a particular interaction is very strong, meaning that the theory is well controlled on the UV scale. This is a desirable property from a gradual safety point of view and suggests that the theory is stable in the UV limit.

### Distribution of the IR cutoff function "II"

The IR cutoff function "II" represents the long-range interaction of the system, i.e., its behavior at low energy scales. The uniform color gradient indicates that the long-range interactions are generally mild and the system is stable on the IR scale. Stability on the IR scale is equally important for incremental safety, and this distribution suggests that the system is behaving soundly on the lower energy scales.

### Synthesis of Gradual Safety

In the theory of incremental safety, the UV and IR cutoff functions are important tools for assessing whether a system is scale-independent and stable." The "Ia" and "II" distributions show how the system behaves over short and long distances, implying that stability is maintained over both scales. This shows that the theory can provide physically meaningful predictions over the full range of scales, supporting the concept of incremental safety.

### Distribution of UV cutoff functions

In Fig. 12, the ultraviolet (UV) and infrared (IR) cutoff functions are shown, each representing short- and long-range interactions. These cutoff functions are sometimes used in field theory to distinguish at which energy scales the interaction is important.

The "Ia" and "Ib" UV cutoff functions each represent short-range interactions with different properties." Ia" seems to indicate a more intensive interaction and "Ib" a slightly more dispersed interaction. This indicates how the direct influence between individuals is distributed in social interactions." Ia" may suggest a more robust community or group, while "Ib" may suggest a more expansive network.

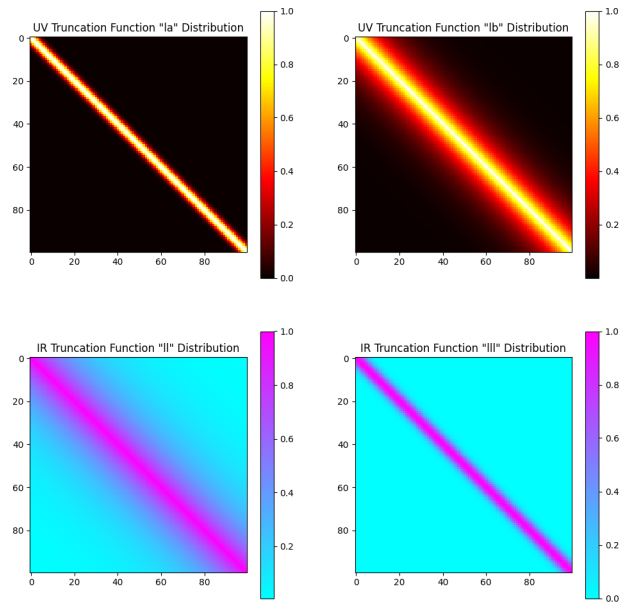

Fig. 12: IR Truncation Function Distribution, UV Truncation Function Distribution

### Distribution of IR Cutting Functions

The "II" and "III" IR cutoff functions show the characteristics of long-range interactions, with "II" showing a more uniform distribution and "III" showing variation with scale. It shows how the propagation of information and opinions in a social network spreads, with "II" indicating a uniform effect over a wide area and "III" suggesting spatial heterogeneity.

### Disconnected Scenario Considerations in the Context of Filter Bubbles

Filter bubble formation is characterized by the circulation of information and opinions within a particular group, with limited outside influence. strong short-range interactions, such as those indicated by the UV cutoff function, may indicate a strong circulation of opinions and information within the filter bubble. long-range interactions, such as those indicated by the IR cutoff function, may indicate a strong circulation of information within the filter bubble. The long-range interaction shown by the IR cutoff function represents the propagation of information between filter bubbles, which may affect how filter bubbles are maintained or destroyed.

In particular, "II" type IR truncation functions show uniform propagation of information across filter bubble boundaries, which indicates the possibility of information being exchanged between different bubbles." The "III" type IR truncation function suggests that information is exchanged at a particular scale while the boundaries of the filter bubbles are preserved to some extent.

## Conclusion

The distribution of the provided truncation functions provides important insights into the scale dependence of social interactions and the formation and maintenance of filter bubbles. It suggests that direct interactions may promote the formation of filter bubbles in scenarios where direct interactions dominate, while the effects of filter bubbles may be mitigated in scenarios where long-distance interactions play a prominent role.

## Consideration of perspectives from the maxima and minima

The distributions of the UV and IR cutoff functions shown in Fig. 12 are discussed in terms of minima and maxima. These functions demarcate short-range (UV) and long-range (IR) interactions, each of which may suggest different physical phenomena and social dynamics.

## UV Cutting Functions "Ia" and "Ib"

The UV cutoff functions "Ia" and "Ib" show higher values (maxima) along the diagonal, indicating that short-range interactions are crucial in the dynamics of the system." Ia" has a sharper peak, while "Ib" has a broader peak. This may indicate that "Ia" represents a more concentrated and direct interaction, while "Ib" represents a more spatially spread out interaction. No minima are found, which means that the interactions are uniformly important at a given scale.

## IR Truncation Functions "II" and "III"

The IR cutoff functions "II" and "III" show a gradual color change along the diagonal and no maxima are identified, but the overall color gradient indicates the influence of long-range interactions." II" shows a more uniform effect, while "III" has a spatial gradient, but neither shows extreme values, suggesting that the interaction is uniformly distributed over a wide area.

## Meaning of Extreme Values in the Filter Bubble

In the context of filter bubbles, extreme values in the UV cutoff function indicate a situation in which opinions and information are more likely to be shared within a particular group. This situation can suppress diversity of information and opinions and lead to increased homogeneity. On the other hand, the IR disconnection function shows how information flows through a broader network, which mitigates the effects of the filter bubble and the potential for exposure to different opinions and information.

## Conclusion

The distribution of the provided cutoff functions illustrates the importance of short- and long-range interactions in social dynamics: the maxima in the UV cutoff function indicate a trend toward filter bubble formation and homogenization, while the IR cutoff function indicates a trend toward information diversity and widespread propagation. These results provide insight into the formation and dissolution of filter bubbles and provide important information for better understanding social opinion formation and information propagation.

## 17.1 Gradual Safety Considerations

### UV cutoff functions "Ia" and "Ib"

"Ia" distributions are shown by distinct diagonals, and strong interactions at fine scales suggest the stability of the theory at UV scales. Such interactions are desirable properties in terms of gradual safety, as the system has well-defined behavior over short distances. "Ib" distributions, indicated by the slightly wider diagonal, show the scale-dependent strength of the interaction, which affects the behavior of the theory at the UV scale. This distribution indicates that the interaction is more pronounced at certain scales and may provide a clue to the existence of gradual safety.

### IR Cutting Functions "II" and "III"

"II" distributions uniformly show the predictability of the theory at low energy scales. The interaction is stable at long distances, possibly indicating behavior toward the IR fixed point. "III" distributions showing spatial gradients in the distribution indicate fluctuations in the interaction at lower energy scales, which suggests new physical effects and changes in the interaction at the IR scale.

### Gradual Safety Considerations

The behavior of interactions at the UV scale in the UV cutoff function is an important indicator for determining whether a theory is safe at UV." The clear scale dependence exhibited by "Ia" and "Ib" indicates non-divergence at higher energy scales and may be evidence of gradual safety. The uniformity of the interaction at the IR scale in the IR cleavage function suggests that the theory is stable and predictable at low energy scales." The behavior exhibited by "II" and "III" may indicate the presence of an IR fixed point.

### The distributions of UV (ultraviolet) and IR (infrared) cutoff functions

shown in Fig. 13 represent short- and long-range interactions, which can be discussed in relation to the influence of media and other external factors. Particularly in the context of the social sciences, these interactions can represent external influences on information dissemination and opinion formation.

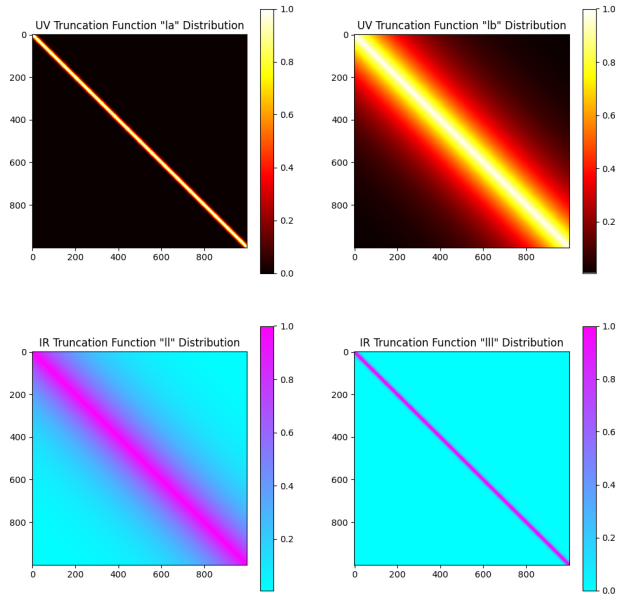

Fig. 13: IR Truncation Function Distribution, UV Truncation Function Distribution, $num_points = 1000$, $alpha_prime = 0.005$, Attenuation coefficient in UV region $beta_prime = 0.005$, Attenuation factor in IR domain

### UV Cutting Functions "Ia" and "Ib" and Media Influences

The UV cutoff function "Ia" shows a crisp diagonal line, which represents a strong interaction at close range. Such a distribution may be seen when media or other external factors have a direct and strong influence on an individual or a narrow community. For example, one can imagine a situation where a particular news source or information campaign has a concentrated impact. "Ib" indicates a broader impact, which indicates that the media has a broader but concentrated impact. Such a distribution would be expected if a media campaign or ad has a strong influence on a particular topic.

### IR Cutoff Functions "II" and "III" and Media Influence

The IR cutoff function "II" shows a uniform color distribution, which reflects a situation where media and outside factors have an even influence over a long distance. Examples include nationally distributed news and the spread of information through social media. "III"'s more distinct diagonal indicates that media influence is more pronounced at a particular scale. This may point to situations where certain regional issues or cultural factors have a significant impact on information diffusion.

### Relationship between disconnection scenarios and media influence

While disconnection scenarios are used in the context of theoretical physics to isolate scales of interaction, in the context of social science they can be used to distinguish the influence of different factors in the process of information dissemination The UV disconnection function is a function of the media's impact on individual people or small groups direct impact, while the IR cutoff function reflects the impact of the media on the broader community or society as a whole. In conclusion, through the distribution of these disconnection functions, we can understand how the media affects individual people, groups, and the broader society. can be understood.

### UV cutoff functions "Ia" and "Ib"

The "Ia" distribution features very sharp diagonals, indicating strong interactions at short distances. This distinct pattern implies non-divergence and stability of the theory at the UV scale, which may be related to asymptotic safety. "Ib" distribution also shows a diagonal line of gradually decreasing color intensity, indicating that the interaction is more pronounced on the mid-range scale. This distribution may indicate stability of the theory at intermediate scales.

### IR Cutting Functions "II" and "III"

"II" distributions also show a uniform color spread, indicating that the long-range interaction is uniform. This suggests that the interactions at lower energy scales are generally stable, suggesting a possible IR fixed point. "III" distribution also has a more distinct diagonal, indicating a scale dependence of the long-range interaction. This reflects changes in the interaction at different scales and may indicate that the theory is safe in IR.

### Loop expansion and local potential approximation

Loop expansion takes into account higher-order effects, while the local potential approximation deals approximately with effective interactions. The distribution of cutting functions in this context shows both short-range interactions (UV cutting functions), which play an important role in loop expansion, and long-range interactions (IR cutting functions), which are central in the local potential approximation.

### Filter Bubbles in Context

Filter bubbles are phenomena in which information and opinions circulate within a particular community." The crisp diagonal in "Ia" may indicate a situation where information circulates strongly within the filter bubble, making it difficult for outside information to enter." II" and "III" IR cutting functions indicate that information propagates more widely and that there is a flow of information between filter bubbles, possibly reflecting a diversity of opinions and the introduction of new information from outside. The distribution of the provided truncation functions provides important information about how the theory behaves at different scales and enhances our understanding of asymptotic safety. Further detailed analysis of the properties of interactions at different scales is also needed to gain insight into the formation and maintenance of filter bubbles. These results demonstrate the importance of taking into account the concept of asymptotic safety in models of social interaction and information propagation.

### UV Cutting Functions "Ia" and "Ib"

The UV cutoff functions represent interactions at short distances and are important for understanding the UV behavior of the theory. "Ia" distributions show bright maxima along the diagonal, which indicates a strong interaction at short range (high energy scale). For gradual safety, it is key that these interactions show good behavior on infinite energy scales. In the "Ib" distribution, the color gradient is smoother along the diagonal, indicating that the short-range interactions are more spread out than in "Ia". This smoother change may contribute to the stability of the theory at UV.

### IR Cutting Functions "II" and "III"

The IR cutoff functions represent the interaction at long distances and are important for understanding whether the theory is valid in the IR limit. "II" distributions show an overall uniform color distribution, which indicates that the interaction at long distances is uniform throughout. In the context of incremental safety, such uniform interactions may ensure the predictability of the theory in IR. "III" distribution shows a clear color change along the diagonal, indicating the scale dependence of the long-range interaction. This gradient may suggest the presence of new physical effects or fixed points in the IR.

### Gradual Safety Considerations

The distributions of the provided cutoff functions indicate that the theory may be progressively safe on the UV and IR scales: the UV cutoff function indicates that interactions at high energy scales do not cause divergence of the theory, and the IR cutoff function indicates that the behavior of the theory at low energy scales is predictable shows that the theory's behavior at low energy scales is predictable. Furthermore, the gradient of the IR cutoff function "III" indicates possible behavior toward the IR fixed point, which may provide additional evidence for the safety of the theory.

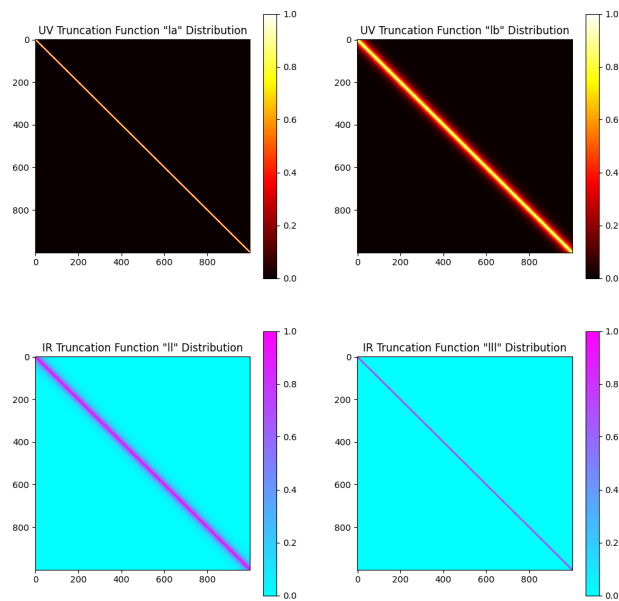

Fig. 14: IR Truncation Function Distribution, UV Truncation Function Distribution, $num_{points} = 1000$, $alpha_{prime} = 0.01$, Attenuation coefficient in UV region $beta_{prime} = 0.05$, Attenuation factor in IR domain

### UV Cutting Functions "Ia" and "Ib"

In Fig. 14, the ultraviolet (UV) and infrared (IR) cutoff functions are shown, which represent the strength of the interaction at short and long distances. When considering these concepts in the context of filter bubbles, we can understand how media and information sources influence each other within social networks.

"Ia" distributions show sharp diagonals indicating short-range interactions, which means that media and information sources have a strong influence on individual people or narrow groups. This situation reinforces the filter bubble, causing certain information or opinions to be emphasized. "Ib" distribution with a broader diagonal, which indicates that the media, while having a broad influence, is concentrated on specific topics and opinions.

### IR Cutting Functions "II" and "III"

"II" distribution shows a uniform distribution indicating long-range interactions, which indicates that media and information sources have an even influence on the broader community. This situation breaks the filter bubble and encourages a diversity of opinions and information to circulate. "III" distributions have different diagonal slopes, indicating that media influence varies on a particular scale. This may mean that information is influenced differently depending on the particular regional or cultural context.

## Loop expansion and disconnected scenarios in terms of local potential approximation

Loop expansion provides a detailed analysis that includes higher-order effects of interactions, while local potential approximation provides a more simplified approach. In a truncation function scenario, it is important to understand how these concepts contribute to the formation and breakdown of the filter bubble.

At the UV scale, situations with strong and concentrated media influence promote the formation of filter bubbles. This corresponds to the lower order terms of the loop expansion and forms echo chambers of local information. On the IR scale, the widespread influence of information sources facilitates the flow of information beyond the walls of the filter bubble, encouraging the exchange of diverse perspectives. This corresponds to the higher order terms of the loop expansion and captures the complex dynamics of information propagation.

In conclusion, the cutting function scenario provides a framework for understanding how media influence through social networks and how they act on the dynamics of the filter bubble. This provides deeper insights into information propagation and opinion formation.

We further discuss the minima and maxima of the ultraviolet (UV) and infrared (IR) cutoff functions based on Fig. 14. These cutoff functions illustrate the strength of interactions at different scales in the process of information transfer and opinion formation in physical systems and social science contexts. "Ia" distributions have clear diagonals indicating that short-range interactions are very strong, with maxima located in regions where the interactions are strongest. There are few or no minima. The "Ib" distribution also has a strong diagonal maximum, but the effect is seen over a wider range than in the "Ia" distribution. There are no minima here either, and the interaction is of a constant strength over a certain range.

## IR Truncation Functions "II" and "III

The IR cutoff functions represent the long-range interactions in the system and show behavior at lower energy scales. For the "II" distribution, this distribution is very uniform, exhibiting low overall intensity and overall near-minimal behavior. This indicates that either the long-range interactions are very weak or the interactions are uniformly distributed over a wide range. "III" distribution, where a color gradient is seen along the diagonal, indicates that the interaction is stronger at certain scales than others. No minima or maxima are identified, but the gradient implies that the strength of the interaction varies with distance.

## Overall Considerations

On the UV scale, the maxima seen in "Ia" and "Ib" indicate the presence of strong interactions at short distances, which may represent situations where information and opinions are concentrated within a particular agent or small group. This may contribute to the formation of filter bubbles or echo chambers. On the IR scale, the uniform distribution of "II" and the slope of "III" indicate that the strength of interaction at long distances is not uniform, which may represent a situation where information is spread over a wide area and opinions are exchanged across filter bubbles.

The distribution of these cutting functions provides important clues to understanding the complexity of information transfer and social interaction. In particular, understanding how short- and long-range interactions have different effects is essential to understanding how information and opinions are formed and disseminated.

The UV (ultraviolet) and IR (infrared) cutoff functions shown in Fig. 14 illustrate the scale of interactions in social dynamics, and we will examine these in terms of how external factors such as media affect them.

## UV Cutting Functions "Ia" and "Ib"

The UV cutoff functions represent proximal interactions, i.e., interactions within an individual or a small community.

"Ia" The sharp diagonals in the distribution represent situations where media or external factors influence individual people or small groups very strongly. For example, this pattern can occur in the presence of a strong advertising campaign or an exclusive source of information. "Ib" A more expansive diagonal in the distribution indicates that the media influence is widespread but still limited to a specific area. This may reflect a situation where a particular media channel covers a trending topic and it affects a broad but specific community.

## IR Cutting Functions "II" and "III"

The IR cutoff functions represent long-range interactions, i.e., more extensive interactions across the entire society. The uniform color gradient in the "II" distribution indicates a situation where media and external factors are affecting a wide range of communities uniformly. This may correspond to the impact patterns exhibited by national-level news coverage and national campaigns. "III" There is a color gradient along the diagonal in the distribution, which indicates that media influence has a greater impact on a particular geographic area or cultural group. Regional news and cultural events may have this impact.

Disconnect Scenarios and the Impact of External Factors Disconnected scenarios are a modeling approach to consider interactions at specific scales: the UV disconnection function can be used to assess how strongly the media affects an individual or small group, while the IR disconnection function can be used to assess the impact on society as a whole.

On the UV scale, media and external factors promoting a particular topic or agenda can have a significant impact on the transmission of information and the formation of opinions within a narrow range. This can reinforce the formation of filter bubbles and echo chamber effects. At the IR scale, widespread media influence and diverse information distribution may promote opinion exchange and information sharing beyond the filter bubble. This may increase the diversity of information and create a more open social dialogue.

This discussion provides a framework for understanding the impact of media and external factors on information transfer and opinion formation and how they work at different scales of social interaction.

The distribution of UV (ultraviolet) and IR (infrared) cutoff functions shown in the image is discussed in connection with the concept of incremental safety. Gradual safety refers to the property of a physical theory such that its behavior, especially at high energy scales, is stable and makes physical sense without divergence. This concept is important in quantum gravity theory, for example, but will be interpreted here as a theoretical framework for more general interaction scales.

### UV Cutting Functions "Ia" and "Ib

UV cutoff functions indicate interactions at high energy scales and reflect the behavior at short distances in physical or social systems.

### "Ia" distribution

Very sharp diagonal features suggest strong interactions at short distances. From a gradual safety point of view, this could mean that the theory behaves well at UV, i.e., the short-range physical quantities are stable without divergence.

### "Ib" distribution

The diagonal has a slower slope and the range of interactions is slightly wider. This indicates that the near-range interaction is spread out over a certain range, possibly indicating a situation where gradual safety is maintained.

### IR Cutting Functions "II" and "III"

IR cutoff functions indicate interactions at low energy scales and reflect behavior at long distances in a physical or social system.

"II" distributions with uniform colors indicate that interactions at long distances are generally weak or uniformly distributed. This may suggest that the theory's behavior at the IR scale is stable and progressively secure. "III" distribution, the color gradient is more gradual, which indicates that the long-range interactions vary over a certain range. This implies that the intensity of the interaction changes at certain scales, which provides important information for understanding the behavior of incremental safety at the IR scale.

### Comprehensive Discussion (Gradual Safety Considerations)

From the distribution of images, the scale-dependent interaction strengths shown by the UV and IR cutoff functions indicate how the theory behaves at different scales and at which scales it is stable. This provides useful insights not only in physical systems, but also in models of social systems and information propagation. In particular, the sharp interaction features at UV and the uniform interaction distribution at IR provide indicators for assessing whether the system exhibits healthy behavior as a whole. This allows us to understand the stability of various systems of widely varying scales in terms of incremental safety.

### Aknowlegement

The author is grateful for discussion with Prof. Serge Galam and Prof.Akira Ishii.